\documentclass[useAMS,usenatbib]{mn2e}
\usepackage{psfig}
\usepackage{times}

\newcommand{\omb}{\Omega_{\rmn{orb}}}

\newcommand{\msol}{\rmn{M_{\sun}}}

\title[Long time-scale variability in GRS~1915$+$105]{Long time-scale variability in GRS~1915$+$105}
\author[M. R. Truss and G. A. Wynn]{M. R. Truss$^{1}$\thanks{E-mail:mrt2@st-andrews.ac.uk} and G. A. Wynn$^{2}$\\
$^{1}$School of Physics \& Astronomy, University of St Andrews, North Haugh, St Andrews, Fife, KY16 9SS, UK\\
$^{2}$Department of Physics \& Astronomy, University of Leicester, University Road, Leicester, LE1 7RH, UK}

\begin{document}

\date{Accepted ~~~ Received ~~~}

\pagerange{\pageref{firstpage}--\pageref{lastpage}} \pubyear{2004}

\maketitle

\label{firstpage}

\begin{abstract}
We present very high resolution hydrodynamical simulations of accretion discs in black hole X-ray binaries accreting near 
the Eddington limit. The results show that mass loss, irradiation and tidal interactions all have a profound effect on the observed behaviour of long 
period X-ray transients. In particular, the interplay of all of these effects in the outer regions of the accretion disc is able to drive long time-scale 
(weeks to years) variability is these objects, and is a possible origin for some of the extreme variability of GRS~1915+105.
\end{abstract}


\section{Introduction}

The outbursts of long period X-ray binaries involve accretion rates close to the Eddington
limit. Consequently, the evolution of the massive accretion discs in these objects is expected 
to be highly influenced by mass loss and irradiation by the central X-ray source. Outbursts of this 
nature are thought to power the most luminous stellar X-ray sources such as the galactic microquasars 
and at least some of the recently discovered ultra-luminous X-ray sources in nearby galaxies \citep{kin02}.
In this paper we investigate the evolution of accretion discs and the origin of the long-term variability 
in the outbursts of long period X-ray transients, using the galactic microquasar GRS~1915+105 as an example system.

\begin{figure}
  \psfig{file=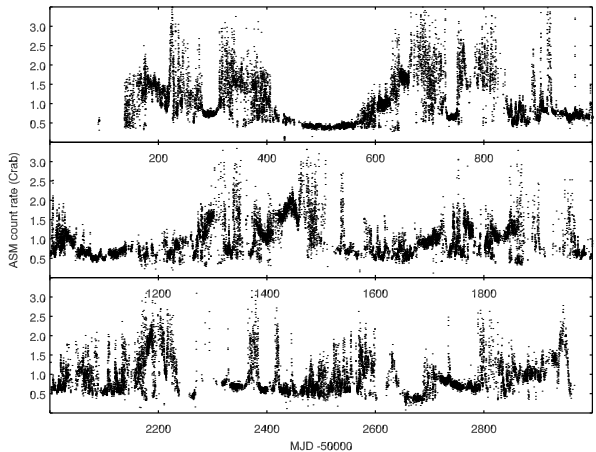,width=8cm}
  \caption{{\it RXTE\/} All-Sky Monitor (ASM) X-ray flux observations of GRS1915$+$105. The system shows a wealth of variability over a wide range 
of time-scales and has been in a continuous state of outburst since 1992. {\it RXTE\/} results are provided by the ASM/{\it RXTE\/} teams at MIT and at 
the {\it RXTE\/} SOF and GOF at the NASA Goddard Space Flight Centre.}
  \label{obs}
\end{figure}

The galactic microquasar GRS~1915+105 (V1487 Aql) was discovered as an X-ray transient in 1992
\citep{cas}, and has been observed to be 
extremely luminous ever since. This binary system contains a 14 $\msol$ black hole \citep{gre01a} accreting from a late-type giant of mass $0.8 \pm 0.5 \msol$ 
\citep{har} via Roche lobe overflow. GRS~1915+105 is unique among 
accreting Galactic black holes spending much of its time at super-Eddington luminosities \citep{don}. 
It is an extremely variable source, exhibiting dramatic, aperiodic variability on a wide range of timescales, from milliseconds to 
months. Figure \ref{obs} shows the Rossi X-ray Timing Explorer All-Sky Monitor observations of X-ray flux from GRS~1915+105
over the past decade.

The fundamental origin of this spectacular behaviour is thought to be an outburst similar to those 
exhibited by other X-ray transients, but of a much longer duration. The outbursts of X-ray transients 
are believed to be initiated by a thermal-viscous accretion disc instability, and prolonged by the heating effect
of the central, irradiating X-ray source \citep{kin98,dub,tru02}. The extremely long time-scale of the outburst of GRS~1915+105 is 
a consequence of its long 33.5 day orbital period, which implies a very large disc size ($R_{\rmn{disc}} \sim 3 \times 10^{12}~\rmn{cm}$). 
Such a large disc can accumulate an enormous amount of mass during quiescence, which is able to support a subsequent super-Eddington
outburst over many decades.

So far, most models of the variability of GRS 1915+105 have concentrated on explaining the origin of the
short time-scale luminosity changes associated with the inner regions of the accretion disc (for example, Belloni et al. 1997), although \citet{kin04} have
proposed a mechanism for generating variability on a wider range of time-scales, in which outflows of mass from the disc are driven by 
localised magnetic dynamo processes . However, the origin of longer-term, aperiodic variations on timescales of hundreds of days remain poorly 
understood as these are most likely driven by dynamical evolution of the outer disc. 

In this paper we present the results of two-dimensional (2D) calculations of the long-term evolution of an accretion disc in a long-period X-ray 
binary accreting close to the Eddington limit, using GRS~1915+105 as an example system. We include the effects of tidal instability, central 
irradiation and mass loss on the disc evolution. We show that the interplay of these effects in the outer regions of the disc can induce similar 
variability to that observed in GRS~1915+105. 

\section{Numerical Model}

In this section we discuss several key refinements that have been made to an existing smooth particle hydrodynamics 
(SPH) code. 

SPH is a Lagrangian scheme used to describe the dynamics of fluid flow. A continuous gas is described by an ensemble 
of particles moving with the local fluid velocity. Local properties such as velocity, density and temperature are 
determined by an interpolation over the neighbouring particles and in this way the fluid equations may be solved. In the 
continuum limit (that is, for a large enough number of neighbour particles), the SPH equations for momentum, thermal 
energy and so on reduce to the exact fluid equations. The reader is referred to \citet{mon} for a review.

The code used for this work is based on a vectorised version of the three-dimensional accretion disc code developed by 
\citet{mur96}. There are four key elements of physics that must be included in a code that attempts to describe the 
behaviour of a system like GRS 1915$+$105, accreting near the Eddington limit: an implementation of the viscous 
instability to drive the outburst, a description of the tidal forces of the secondary star on the disc, the effect of 
self-irradiation of the disc by the X-rays generated near the primary and mass loss when the Eddington limit is breached.
We will describe each of these in turn.

\subsection{Viscous instability}

The disc instability, which is driven by the onset of ionization in the temperature range 6,000 to 7,000 K, is implemented 
by a viscous switch based on the local disc conditions. The viscosity, $\nu$, is given by the formula of \citet{sha}:
\begin{equation}
   \nu = \alpha c_{\rmn{s}} H,
\end{equation}
where $c_{\rmn s}$ is the local sound speed, $H$ the scale height and $\alpha$ is the viscosity parameter. It is well
known that the increase in temperature at the transition from quiescence to outburst is insufficient to provide the
angular momentum transport required to drive the mass accretion rates observed during an outburst. Hence, it is
inferred that the heating is accompanied by an increase in the viscosity parameter $\alpha$. In previous SPH studies of
outbursts in accretion discs, $\alpha$ was allowed to change instantaneously \citep{mur98}, or on some fixed time-scale 
\citep{tru00}, but $c_{\rmn{s}}$ was kept fixed. Here, we allow the sound speed as well as the viscosity parameter to 
vary on a time-scale that is determined by the local conditions. Recently, a similar method was applied to the outbursts of 
cataclysmic variables \citep{tru04}. The method is able to reproduce a full, realistic outburst amplitude with none of 
the scaling factors required by the previous schemes that used a constant sound speed everywhere. In this approach, we 
would not reproduce the expected temperature profile of a disc in steady state, where $c_{\rmn{s}} \propto R^{-3/8}$, because we
have lower and upper bounds to the sound speed (but allow a continuum of states between). However, as we shall see, once the outburst is underway 
the disc as a whole spends little or none of its time in a single state and most of its behaviour is governed by local processes; changes in 
sound speed and mass loss occur in relatively small regions, rather than globally across the disc. It is not clear, then, that a fully self-consistent
calculation - computationally, this remains inviable in a two-dimensional hydrodynamic simulation of this magnitude - would produce vastly different 
results. Until this becomes possible, we believe that the method we describe here, which allows for realistic local changes in sound speed, offers a
better prospect than the fully-isothermal schemes of the past.

The sound speed and temperature are changed in response to the local surface density calculated at the position of
each particle. We define critical surface densities, $\Sigma_{\rmn{max}}$ and $\Sigma_{\rmn{min}}$ to trigger an outburst
and to return to quiescence respectively. These critical values scale almost linearly with radius in the disc \citep{lud}. When the surface
density at a particle crosses one of these values, both the sound speed and the viscosity parameter are changed 
smoothly according to
\begin{equation}
   c_{\rmn{s}}(t)= \frac{(c_{\rmn{s,hot}}+c_{\rmn{s,cold}})}{2} \pm \frac{(c_{\rmn{s,hot}}-c_{\rmn{s,cold}})}{2} \tanh 
   \left( \frac{t}{t_{\rmn{th}}} - \pi \right)
\end{equation}
and
\begin{equation}
   \alpha(t)= \frac{(\alpha_{\rmn{hot}}+\alpha_{\rmn{cold}})}{2} \pm \frac{(\alpha_{\rmn{hot}}-\alpha_{\rmn{cold}})}{2} 
   \tanh   \left(\frac{t}{t_{\rmn{th}}} - \pi \right).
\end{equation}
The thermal time-scale for the change, $t_{\rmn{th}}$, is defined as the ratio of the heat content per unit (projected) area of the disc, $\Sigma 
c_{\rmn{s}}^{2}$, to the total rate of energy loss per unit area, which taking into account both faces of the disc is  $9 \nu \Sigma GM_{1} / 4R^{3}$.
Therefore, the thermal time-scale
\begin{equation}
   t_{\rmn{th}} = \frac{4}{9 \alpha \Omega}.
\end{equation}

\subsection{Tidal forces and mass transfer}

The calculations are performed in the full Roche potential of a binary system. In this way the tidal influence of the 
secondary star is intrinsic to the scheme. For the mass ratio $q = M_2 / M_1 \sim 0.07$ inferred for GRS~1915$+$105, we expect the 
tides to have a significant effect on the orbits near the outer edge of the accretion disc, and to launch a two-armed spiral wave. Mass 
transfer from the secondary star is implemented by the injection of particles at the inner Lagrangian point. The rate of mass injection is 
independent of the physical conditions at the edge of the disc and is kept constant throughout the calculation, regardless 
of whether the disc is quiescent or in outburst.

\subsection{Irradiation}

We use a simple model due to \citet{kin98}, in which the X-rays emitted near the black hole keep the disc in the hot, high 
viscosity state out to a certain radius $R_{\rmn{h}}$ which is determined by the central accretion rate. Matching the
irradiation temperature to the minimum temperature to keep hydrogen ionised, $T_{\rmn{H}}$, we have
\begin{equation}
   R_{\rmn{h}}^{2} = \frac{\eta \dot M c^{2}}{4 \pi \sigma T_{\rmn{H}}^{4}} f\left( \beta, H, R \right)
\end{equation}
where $\eta$ is the accretion efficiency and $f$ is a factor dependent on the albedo, the disc geometry and the form of 
the irradiating source. For the typical parameters of a disc around a black hole given in \citet{kin98}, we have
\begin{equation} 
    R_{\rmn{h}} = 1.9 \times 10^{11} \left( \frac{\eta}{0.1} \right)^{\frac{1}{2}} \dot M_{\rmn{8}}^{\frac{1}{2}} \,\, \rmn{cm,}
\end{equation}
where $ \dot M_{\rmn{8}}$ is the central accretion rate in units of $10^{-8} \, \msol \rmn{yr^{-1}}$.
In the calculations, all particles within $R_{\rmn{h}}$ are given the high state viscosity $\alpha_{\rmn{hot}}$ and sound
speed $c_{\rmn{s,hot}}$.

\subsection{Mass loss}

In the next section we will discuss the results of three calculations of outbursts in GRS1915$+$105. In each calculation, 
the mass loss due to the Eddington limit being exceeded is dealt with in a different way. In the first, which is intended to
act as a control case, the limit is ignored and $R_{\rmn{h}}$ is allowed to evolve for accretion rates beyond the 
Eddington limit. In the second, $R_{\rmn{h}}$ is only allowed to increase up to a maximum value corresponding to a 
central accretion rate equal to the Eddington rate. 

In the third calculation, individual particles are removed locally from the accretion disc where the Eddington limit is 
breached, not just at the inner boundary. The local mass accretion rate at each radius in the disc is calculated from
\begin{equation}
  \dot M(r) = -2 \pi r v_{\rmn{r}}(r) \Sigma(r).
\end{equation}
The Eddington accretion rate at a radius $r$ is
\begin{equation}
   \dot M_{\rmn{Edd}}(r) = \frac{r L_{\rmn{Edd}}}{GM_{1}}
\end{equation}
where $L_{\rmn{Edd}}$ is the Eddington luminosity. Particles are removed from the disc before they cross the inner disc
boundary wherever $\dot M(r) > \dot M_{\rmn{Edd}}(r)$, or
\begin{equation}
   \dot M(r) > \dot M_{\rmn{Edd}}(R_{0}) \left( \frac{r}{R_{0}} \right)
\label{loss}
\end{equation}
where $R_{0}$ is the radius of the inner boundary.

It is of primary concern in such a scheme, where particles or groups of particles are being removed from the flow, that 
spatial resolution is not degraded severely. With this in mind, we use a smoothing length that varies according to the
local density. The response of the smoothing length is such that each particle maintains at least 50 nearest neighbour
particles. 

\section{Results}

In this section we present the results of three very large two-dimensional SPH calculations of accretion discs in a
binary system with parameters similar to those of GRS~1915$+$105: $P_{\rmn{orb}} = 33.5~\rmn{days}$, $M_{1} = 14 \, \msol$ 
and $M_{2} = 1 \, \msol$. We use $c_{\rmn{s,cold}} = 0.05 a \omb$, $c_{\rmn{s,hot}} = 0.15 a \omb$, which correspond to
mid-plane temperatures of $\sim 5000~\rmn{K}$ and $\sim 45000~\rmn{K}$ respectively, assuming a mean molecular weight $\mu = 0.6$,
appropriate for a fully ionised mixture of gases (strictly, of course, this value of $\mu$ is not self-consistent for the neutral gas in the low state.  A
better estimate of $\mu$ in the low state would be higher, although this would have little effect on the global evolution of the disc through an outburst 
in the simulations).

 The accretion efficiency is set to $\eta = 0.1$. Our choice of viscosity parameters, $\alpha_{\rmn{cold}} = 0.1$ 
and $\alpha_{\rmn{hot}} = 1$, is slightly higher than we would ideally choose, but is necessitated by our need to simulate a very long time series of data with a 
limited amount of dedicated supercomputer time. We discuss the implications of using different viscosity parameters fully in Section 4. 

The mass transfer rate is calculated from the formula of \citet{rit}:
\begin{equation}
   -\dot M_{2} \simeq 7.3 \times 10^{-10} \left( {M_{2} \over \msol} \right) ^{1.74} P_{\rmn{orb}}^{0.98} \rmn{(d)} \, \,\msol 
   {\rmn{yr^{-1}}},
\end{equation}
which is appropriate for nuclear time-scale-driven mass transfer from an evolved companion star. In this case, the 
formula gives $ -\dot M_{2} \simeq 2 \times 10^{-8} \msol \rmn{yr^{-1}}$. This is already close to the Eddington rate 
(near the black hole); the implications of such a rate are also discussed in Section 4.

Each of the three simulations have identical initial conditions: a steady-state accretion disc with $N_{0} = 439,603$ 
particles. This places these simulations among the largest accretion disc simulations ever attempted with SPH. The simulations were
run on the UK Astrophysical Fluids Facility (UKAFF) during allocated time awarded in late 2003.

\subsection{Simulation 1}

\begin{figure}
   \psfig{file=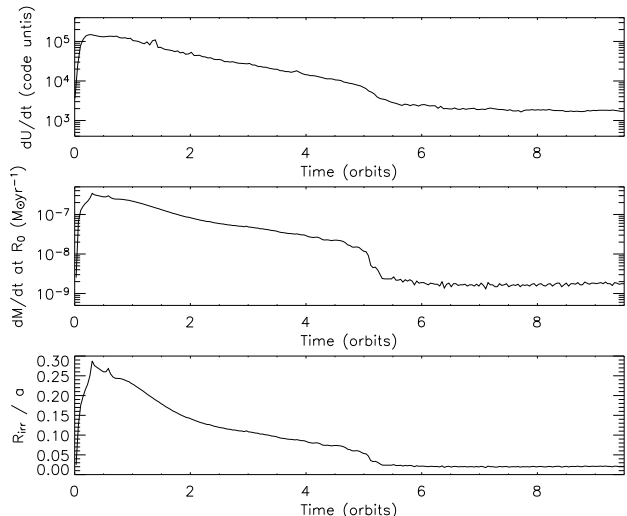,width=9cm}
   \caption{Top to bottom: energy dissipation rate, central accretion rate and irradiated radius as a fraction of binary 
	separation during simulation 1.}
   \label{sim1}
\end{figure}

The accretion disc was allowed to evolve through an outburst with no restriction on the irradiation radius and no mass 
loss. Figure \ref{sim1} shows the resulting time-variations in total rate of energy dissipation per unit mass in the disc (summed over all particles), 
$du/dt$, central accretion rate and irradiated radius, $R_{\rmn{irr}}$.

The central accretion rate rises by two orders of magnitude 
to a peak at $3.5 \times 10^{-7} \rmn{\msol yr^{-1}}$, and the entire disc is irradiated. The outburst profile that results is 
very simple: as the disc is kept in the hot, highly viscous state by the central X-rays, we see an exponential decay as the 
classic 'flat-top' to the outburst. After six months or so, the irradiation radius has retreated to the inner edge of the disc, 
and we see a rapid decline to the quiescent state, which persists until the end of simulation. The number of particles remaining at the
end of the outburst is 189,307, corresponding to the net accretion of $57 \%$ of the total mass of the disc. Each particle has a mass 
$2.09 \times 10^{-13} \rmn{\msol}$, so the total mass accreted during the outburst is $5.2 \times 10^{-8} \rmn{\msol}$.

Clearly, this simulation cannot explain GRS~1915$+$105, but the behaviour 
is consistent with the results of simulations of the short period X-ray transient A0620-003 in \citet{tru02}, where a similarly simple 
outburst was obtained in the case where the entire disc becomes irradiated by the central X-ray flux.

\subsection{Simulation 2}

\begin{figure}
   \psfig{file=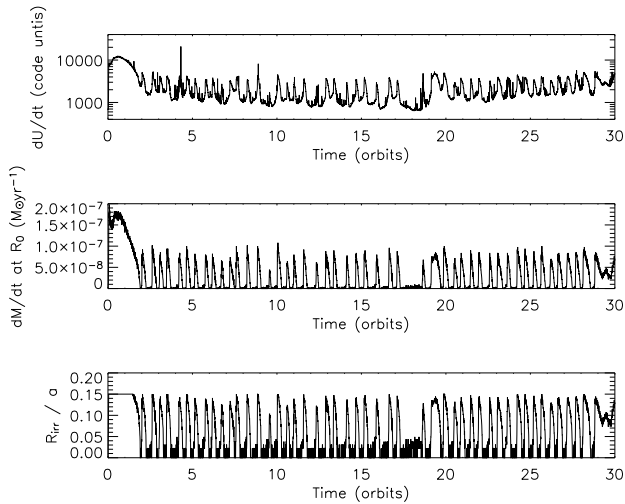,width=9cm}
   \caption{Top to bottom: energy dissipation rate, central accretion rate and irradiated radius as a fraction of binary 
	separation during simulation 2. The brief gaps in the data in the middle panel correspond to times when the 
	central accretion rate was zero.}
   \label{sim2}
\end{figure}

In this simulation, we assume that the Eddington limit holds at the inner boundary of the accretion disc. Material accreting at
a rate higher than $\dot M_{\rmn{Edd}}(R_{\rmn{0}})$ is assumed to be lost from the system and does not contribute to 
increasing the fraction of the disc that is illuminated by the central X-rays. 

Therefore, the irradiated fraction of the disc is 
limited to a maximum value corresponding to accretion at the Eddington rate through the inner boundary. We use 
$\dot M_{\rmn{Edd}}(R_{\rmn{0}}) = 10^{-7} \msol\,\rmn{yr}^{-1}$, which gives $R_{\rmn{h,max}} = 0.15a$, where $a$ is 
the binary separation.

The contrast between the resultant outburst, shown in Figure \ref{sim2}, with that of the previous simulation is stark. As
expected with identical initial conditions, the initial rise to outburst maximum proceeds exactly as before. However, after a
short time, $R_{\rmn{h}}$ reaches its maximum value, and remains there for about two orbits of the binary while the 
central accretion rate is super-Eddington. Following this plateau, the accretion rate declines and enters a phase of aperiodic
variability that is reflected in the energy dissipation rate in the disc. We note from the lower two panels of Figure \ref{sim2}
that the maximum central accretion rate during these flares does not exceed the Eddington rate at the inner boundary. The
flaring behaviour continues until the end of the simulation at $t = 80$ orbits, long after the last data point shown in the figure.
Of course, we do not know for sure if the system would return to quiescence if the simulation could be run indefinitely. 
However, from the data we have, it seems that the system has found a quasi-steady state: the time-averaged accretion 
rate over the whole simulation is very close to  $2 \times 10^{-8} \msol \rmn{yr^{-1}}$, which is the rate at which mass is
added to the disc from the secondary star.

\begin{figure}
~~~~~~~~\psfig{file=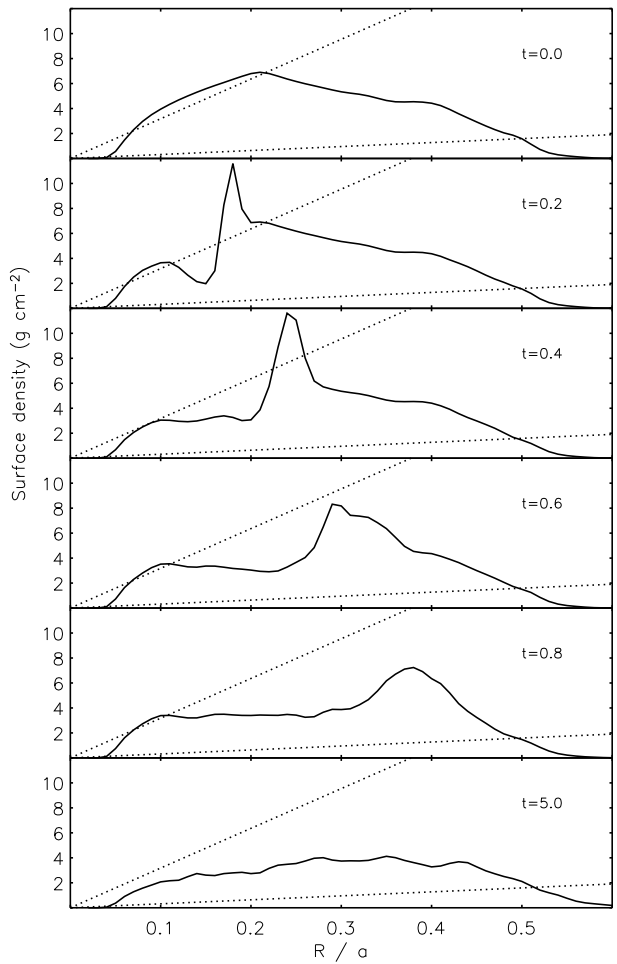,width=8cm}
   \caption{Azimuthally averaged surface density profile during the initial stages of simulation 2. The dotted lines are the 
	critical surface densities $\Sigma_{\rmn{min}}$ and $\Sigma_{\rmn{max}}$ described in the text, although in all but
	the top and bottom panels everything inside $R = 0.15a$ is kept in the hot state by the irradiation. The times are
	quoted in orbital periods.}
   \label{dens}
\end{figure}

Figure \ref{dens} shows the evolution of the surface density profile of the disc during the initial stages of the simulation. 
While the central accretion rate is super-Eddington, a large spike of gas builds up at the hot/cold boundary. As the highly
viscous, hot gas spreads, most of the gas outside $R_{\rmn{h}}$ is pushed over the critical limit $\Sigma_{\rmn{max}}$ and
starts to participate in the outburst. However, this plot does not seem to explain the origin of the flaring behaviour. The 
surface density profiles are produced by averaging over azimuth, and much information about the structure of the flow is lost
in producing them.
\begin{figure*}
~~~~~~~~~~~~~~~~\psfig{file=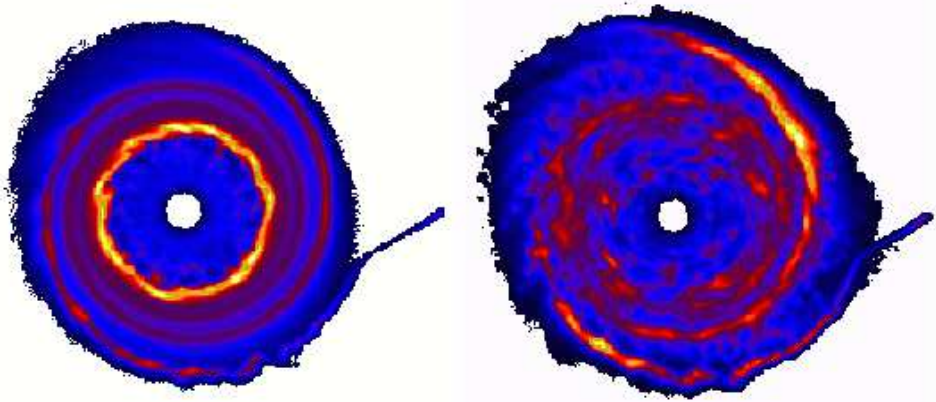,width=14cm,angle=-90}
   \caption{Surface density maps of the accretion disc in simulation 2 at $t$=0.4 orbits (left) and $t$=5 orbits (right). The 
	colour scale runs from black at 0 $\rmn{gcm^{-2}}$ through red to yellow at 12 $\rmn{gcm^{-2}}$. The secondary 
	star to the right of the mass stream and the inner part of the accretion disc(white) are not modelled. The 
	tidally-induced spiral structure and non-circular outer orbits are preserved throughout the simulation despite the 
	inhomogeneity of the flow.}
   \label{discs}
\end{figure*}
\noindent Figure \ref{discs} shows the full density structure of the flow at two times during the simulation; at maximum
$R_{\rmn{h}}$ at $t$=0.4, and later on during the flaring behaviour at $t$=5.0. The initial state of the disc is rather smooth 
with some weak spiral arm features (these can still be seen in the outer part of the disc in the left-hand image of Figure 
\ref{discs}). After the onset of the outburst, however, it can be seen immediately that the flow through the fixed hot/cold 
boundary is neither steady nor smooth. The disc is left with a clumpy, inhomogeneous structure, and it is the accretion of
the over-dense clumps of gas that lead to the subsequent flaring behaviour. This is a direct consequence of fixing 
$R_{\rmn{h}}$ for a short time when the accretion rate onto the black hole is super-Eddington. We have already seen in 
simulation 1 that this structure does not develop if $R_{\rmn{h}}$ is not fixed.

It is interesting to note that the tidally-induced features are preserved as the outer edge of the disc expands 
during the outburst. Despite the unevenness of the flow, a strong two-armed spiral wave is evident by $t$=5.0 orbits
(right hand side of Figure \ref{discs}) that penetrates deep into the disc. In fact, it is the interaction of the spiral arms with the
hot/cold boundary that leads to the variability in this simulation. While the gross structure of the spiral arms is preserved, locally
their structure is determined by the fluctuations in pressure and density induced at $R_{\rmn{h}}$.

The gas near the outer edge moves on eccentric orbits, and this eccentric structure does indeed precess in a prograde fashion in 
exactly the manner expected for a tidally-unstable disc. There is no evidence of superhumps: the periodic variations in dissipation rate observed
in tidally-unstable cataclysmic variables that are caused by gas in orbits resonant with the orbital frequency of the binary. However, 
this is unsurprising because here the dominant component of the energy dissipation rate comes from the irradiated inner disc, not the tidally
stressed outer orbits.

\begin{figure}
~~~~~~~~\psfig{file=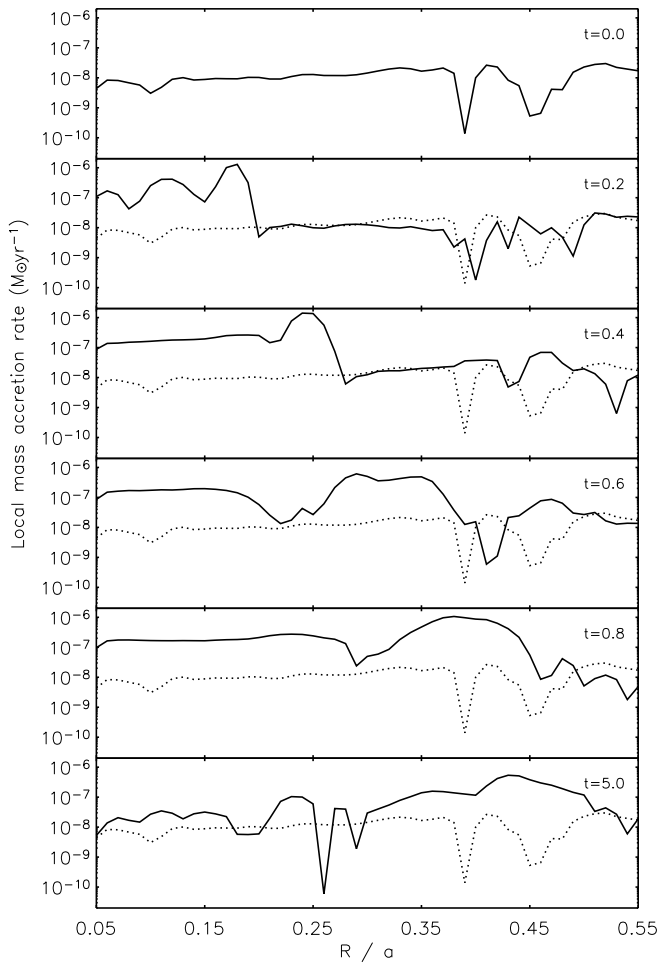,width=8cm}
   \caption{Azimuthally averaged local mass accretion rate during the initial stages of simulation 2. The initial profile is 
	reproduced as a dotted line in each case to aid comparison. The times are identical to those in Figure \ref{dens}.}
   \label{mdot}
\end{figure}

The local mass accretion rate through the disc during the outburst is extremely high. Figure \ref{mdot} shows the evolution 
of the local accretion rate profile for the same time slices as the surface density plots in Figure \ref{dens}. The initial 
steady-state profile is given in the top panel and is reproduced as a dotted line in the lower panels to aid comparison. The 
mass transfer rate at the outer edge of the disc remains constant at all times and save for some structure associated with 
the spiral arms near $R$=0.4$a$, the steady-state profile is constant with radius at a value close to $2 \times 10^{-8} \msol\,\rmn{yr}^{-1}$ 
- the rate of mass transferred from the donor star. When the spike of gas develops
at the irradiated boundary, the local accretion rate rises to a maximum near $10^{-6} \msol\,\rmn{yr}^{-1}$, well sufficient
to sustain a super-Eddington central rate. By $t$=5 orbits, when the main part of the outburst is over and the flaring 
behaviour is underway, the profile approaches that of the steady initial state, but varies in response to the local
inhomogeneity of the flow caused by the clumps and spiral structure.

\subsection{Simulation 3}

\begin{figure}
   \psfig{file=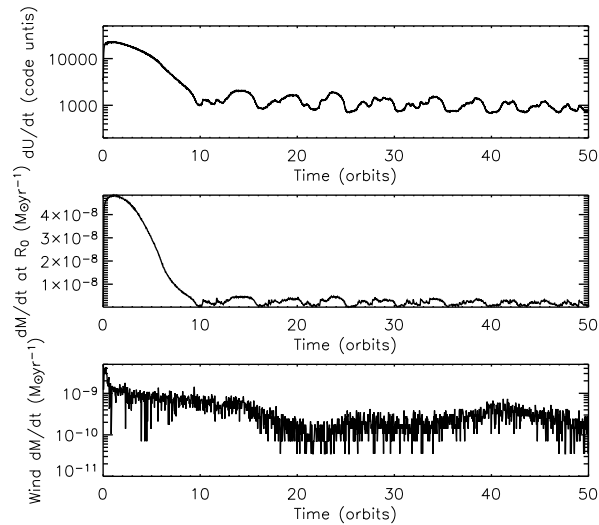,width=9cm}
   \caption{Top to bottom:  energy dissipation rate, central accretion rate and mass-loss rate during simulation 3.}
   \label{sim3}
\end{figure}

\begin{figure}
  ~~~~~~~~\psfig{file=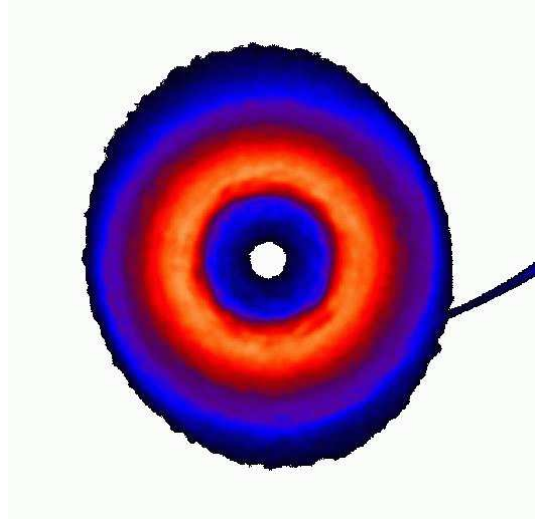,width=7cm}
  \caption{Surface density map of the accretion disc in simulation 3 at t=16 orbits. The colour density scale is the same as for Figure \ref{discs}.
The inner part of the disc is clearly depleted where the wind loss is greatest.}
  \label{disc3}
\end{figure}

\begin{figure}
   \psfig{file=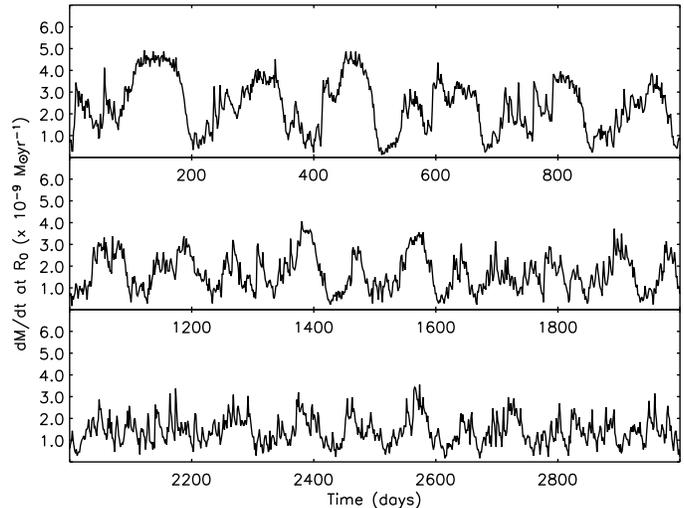,width=9cm}
   \caption{Expanded view of the centre panel of Figure \ref{sim3}, showing mass accretion rate through the inner boundary in simulation 3,
beginning at orbit 10. The data is presented here with time in days (not orbital periods as in previous figures) and has been divided into three
1000-day slices to aid comparison with Figure \ref{obs}.}
   \label{sim3lin}
\end{figure}

Here, mass is removed from the disc locally whenever the accretion rate exceeds the local Eddington accretion rate. This is
the condition that has been defined in equation \ref{loss}, where we use $R_{\rmn{0}} = 0.05a$ and 
$\dot M_{\rmn{Edd}}(R_{\rmn{0}}) = 10^{-7} \msol\,\rmn{yr}^{-1}$ as before. No restriction on $R_{\rmn{h}}$ is necessary, 
because by the time the gas reaches the inner boundary, the accretion rate is by definition less than or equal to the 
Eddington rate. The results are shown in Figure \ref{sim3}. Again, the outburst starts in the same way as 
before, but here, when the large spike in surface density and local accretion rate develops at $R = 0.15a$, a large amount of 
gas is blown away. Unlike in the previous simulations, this gas never reaches the inner boundary, and although variability in 
the central accretion rate is produced, there is less mass involved and the appearance of the outburst profile is simplified. 
This is reflected in the depleted appearance of the disc, especially near the inner boundary, at times later than $t=10$ orbits 
(Figure \ref{disc3}). 
The flares appear to be more regular than in the previous case, although there is some evolution in the variability. This is in response to the large 
mass-loss rate at the beginning of the simulation. Figure \ref{sim3lin} shows an expanded section of 
the mass accretion rate through the inner boundary in three 1000-day sections.  The periodicity of the variations decreases gradually, but by 
$t = 1600~\rmn{days}$ the flares have settles into a quasi-steady state, recurring on a time-scale 
of $\sim 80 - 100~\rmn{days}$. It should be noted that the time-scale of these flares is inextricably linked to the amount of mass lost in the disc, and 
this depends on the value of $R_{\rmn{0}}$ and the accretion efficiency. We discuss this in more detail in Section 4 below. In order to minimise noise,
the central accretion rate is calculated over intervals of 50 time-steps during the simulation, rather than considering the highly discretized case of 
counting an integer number of particles accreted during a single finite time-step. 50 time-steps correspond to 1.3 days in Figure \ref{sim3lin}, hence 
we are confident that even the shorter time-scale variations that we find in the simulations are not attributable to noise.

The bottom panel of Figure \ref{sim3} shows the rate at which mass is removed from the disc before it reaches the black hole. After the initial burst, 
the mass loss rate settles down and oscillates around a mean value of $-\dot M \sim 2 \times 10^{-10} \msol \rmn{yr}^{-1}$.

\section{Discussion}

Our results show that mass loss, irradiation and tidal interactions all have a profound effect on the observable properties of 
long period X-ray transients. In particular, the interplay of all of these effects in the outer regions of the 
accretion disc is able to drive long time-scale (i.e. months to years) variability is these objects, and is a possible origin
for some of the extreme variability of GRS~1915+105.

The physical process that drives these variations occurs in three main stages. In the first stage, after the onset of an outburst, the mass
accretion rate rises and the disc becomes irradiated out to a certain radius $R_{\rmn{h}}$. If the Eddington limit holds, this radius is limited to 
a maximum value corresponding to accretion at the Eddington rate at the inner boundary of the disc. We have shown that the flow through a 
fixed hot/cold boundary is neither steady nor smooth. Gas in the unirradiated part of the disc rapidly crosses the critical threshold for the disc
instability and the remainder of the accretion disc now makes the transition to the hot, high viscosity state. In the third and final stage, the 
interaction of non-axisymmetric structure (here, the spiral waves induced by the tidal stresses of the companion star on the disc) with the
unsteady hot/cold boundary produces a rather inhomogeneous flow. The local variations in density and mass accretion rate produce the variations
in accretion rate seen in our models. A further layer of complexity in the variations is added when local mass loss from the disc (here assumed to be 
in the form of a wind) is taken into account. 

The variable which has the most impact on the variability that is predicted by this model is the magnitude of $R_{\rmn{h,max}}$, the radius to
which the disc can be irradiated for accretion at the Eddington limit. This is sensitive to the geometry of the disc and the illuminating X-ray source 
and to the accretion efficiency. It is probable that self-shadowing and radiative warping of the disc will play some part in the behaviour, which we
have not considered here. If the accretion efficiency is particularly low, $R_{\rmn{h,max}}$ is correspondingly closer to the black hole, perhaps at 
some radius smaller than we can resolve for the inner boundary of our disc in these simulations. While this will impact upon the time-scale of the
variations caused by local mass loss - the rate of mass loss depends on the local Eddington rate, which increases with radius - any variations 
caused by mass loss at smaller radii will occur on shorter time-scales and not affect the gross long-term behaviour that we find in simulation 3. 
Furthermore, the inhomogeneity that starts to appear near the hot/cold boundary rapidly spreads to the whole disc outside $R_{\rmn{h,max}}$. The 
triggering factor for the long time-scale variability is that the  $R_{\rmn{h,max}}$ boundary remains fixed, regardless of its magnitude.

A related issue is the duration of the initial high central accretion rates, which we find at the beginning of simulations 2 and 3. Whether such an
event is present in the observed X-ray light curve in Figure 1 is debatable: the case rests on whether the initial high state that decays over the
first 100 days or so of the outburst is part of a single, coherent event or not. We can, however, comment upon the mechanism that produces the
initial high state in the simulations. It is caused by the accretion of material that starts off inside $R_{\rmn{h,max}}$, and becomes irradiated. It 
follows that the duration of this initial burst will depend on the viscous time-scale at the position  of $R_{\rmn{h,max}}$, which in turn depends on the 
factors discussed above. Here, our choice of accretion efficiency, $\eta$ and $\dot M_{\rmn{Edd}}(R_{\rmn{0}})$ produce a value of $R_{\rmn{h,max}}$ 
that is well outside the inner boundary of the simulated disc. If nature conspires to make  $R_{\rmn{h,max}}$ smaller than this estimate, the 
duration of the initial burst would be shorter.

The postulate that the local wind-loss mechanism must occur at small radii and produce short time-scale variations is indeed supported by the
results of simulation 3. The mean wind loss rate in the simulation from radii greater than the radius of the inner boundary ($R_{\rmn{wind}} > 3.75 
\times 10^{11} \rmn{cm}$) is $\sim 2 \times 10^{-10} \msol \rmn{yr^{-1}}$. This is far less than the estimate of $10^{-6}$ to $10^{-7} 
\msol \rmn{yr^{-1}}$ made by \citet{kot} to explain the hydrogen column density inferred from absorption line spectra. It is more likely that this kind
of mass loss rate is generated at much smaller radii in the disc, where $\dot M_{\rmn{Edd}}$ is much lower.

A fully realistic model for the viscous processes at work in these discs requires non-ideal magnetohydrodynamics (MHD), as the equivalent value 
of $\alpha$ is likely to vary with disc radius, vertical displacement from the mid-plane and with time. While for numerical reasons the values of 
$\alpha$ which we have used in this work are a little higher than are usually assumed for these discs, the discrepancy is not marked. The 
entire disc spends almost all of its time in the high viscosity state during an outburst. MHD calculations have shown that the magneto-rotational 
instability can sustain a mean equivalent viscosity parameter of $\alpha \sim 0.4$ \citep{tou}. While we use $\alpha = 1$ here, the viscosity is 
dominated by the sound speed, scaling as $\nu \propto \alpha c_{\rmn{s}}^2$. The slight overestimate in $\alpha$ is easily compensated by our 
rather conservative estimate of sound-speed in the hot state (described at the beginning of Section 3).

We close by commenting on other factors that contribute to such incredibly high accretion rates during an outburst. If, during a long quiescent period the 
inner regions of the accretion disc are absent through evaporation, the accretion disc may build up a very large reservoir of mass. This could lead to 
extremely high accretion rates during a subsequent outburst. In the case of GRS1915$+$105, however, we expect an unusually high mass transfer rate
from the companion star, and this could be a fundamental factor in the ability of such systems to achieve and maintain accretion rates near, and perhaps
beyond, the Eddington limit. Mass-transfer rates of the order $\sim 10^{-8} \msol \rmn{yr^{-1}}$ are only achievable in low-mass X-ray binaries that 
contain significantly evolved companion stars, transferring mass on a time-scale governed by nuclear expansion of the secondary. It is no surprise, then,
that GRS~1915$+$105, with its long orbital period and huge accretion disc is one of the few systems in which such behaviour has been observed.

\section*{Acknowledgments}
MRT acknowledges a PPARC postdoctoral fellowship. The simulations were performed on the UK Astrophysical Fluids 
Facility (UKAFF). The authors are grateful to Chris Done and also to the referee for helpful and insightful comments. The {\it RXTE\/} light 
curve was provided by the {\it RXTE\/}/ASM team at MIT and {\it RXTE\/} SOF and GOF at NASA Goddard SFC.

\label{lastpage}


\begin{thebibliography}{99}

\bibitem[\protect\citeauthoryear{Belloni et al.}{1997}]{bel} Belloni T., Mendez M., King A.R., van der Klis M., van Paradijs J., 1997, ApJ, 488, L109
\bibitem[\protect\citeauthoryear{Castro-Tirado, Brandt \& Lundt}{1992}]{cas} Castro-Tirado A.J., Brandt S., Lundt N., 1992, IAUC, 5590, 2
\bibitem[\protect\citeauthoryear{Done, Wardzinski \& Gierlinski}{2004}]{don} Done C., Wardzinski G., Gierlinski, M., 2004, MNRAS, 349, 393
\bibitem[\protect\citeauthoryear{Dubus, Hameury \& Lasota}{2001}]{dub} Dubus G., Hameury J.-M., Lasota J.-P., 2001, A\&A,
    373, 251
\bibitem[\protect\citeauthoryear{Ebisuzaki et al.}{2001}]{ebi} Ebisuzaki  T. et al., 2001, ApJ, 562, L19
\bibitem[\protect\citeauthoryear{Frank, King \& Raine}{2002}]{fra} Frank J., King A.R., Raine D.J., 2002, Accretion Power 
in Astrophysics, 3rd edn. Cambridge University Press
\bibitem[\protect\citeauthoryear{Greiner, Cuby \& McCaughrean}{2001}]{gre01a} Greiner J., Cuby J.G., McCaughrean M.J., 
2001, Nature, 414, 422
\bibitem[\protect\citeauthoryear{Harlaftis \& Greiner}{2004}]{har} Harlaftis E.T., Greiner J., 2004, A\&A, 414, L13
\bibitem[\protect\citeauthoryear{King}{2002}]{kin02} King A.R., 2002, MNRAS, 335, L13
\bibitem[\protect\citeauthoryear{King \& Ritter}{1998}]{kin98} King A.R., Ritter H., 1998, MNRAS, 293, L42
\bibitem[\protect\citeauthoryear{King et al.}{2004}]{kin04} King A.R., Pringle J.E., West R.G., Livio M., 2004, MNRAS, 348, 111
\bibitem[\protect\citeauthoryear{Kotani et al.}{2000}]{kot} Kotani T., Ebisawa K., Dotani T., Inoue H., Nagase F., Tanaka Y., Ueda Y., 2000, ApJ, 539, 413
\bibitem[\protect\citeauthoryear{Ludwig, Meyer-Hofmeister \& Ritter}{1994}]{lud} Ludwig K., Meyer-Hofmeister E., Ritter H., 1994, A\&A, 290, 473
\bibitem[\protect\citeauthoryear{Monaghan}{1992}]{mon} Monaghan J.J., 1992, ARA\&A, 30, 543
\bibitem[\protect\citeauthoryear{Murray}{1996}]{mur96} Murray J.R., 1996, MNRAS, 279, 402
\bibitem[\protect\citeauthoryear{Murray}{1998}]{mur98} Murray J.R., 1998, MNRAS, 297, 323
\bibitem[\protect\citeauthoryear{Ritter}{1999}]{rit} Ritter H., 1999, MNRAS, 309,360
\bibitem[\protect\citeauthoryear{Shakura \& Sunyaev}{1973}]{sha} Shakura N.I., Sunyaev R.A., 1973, A\&A 24, 337
\bibitem[\protect\citeauthoryear{Tout \& Pringle}{1992}]{tou} Tout C.A., Pringle J.E., 1992, MNRAS, 259, 604
\bibitem[\protect\citeauthoryear{Truss et al.}{2000}]{tru00} Truss M.R., Murray J.R., Wynn G.A., Edgar R.G., 2000, 
MNRAS, 319, 467
\bibitem[\protect\citeauthoryear{Truss et al.}{2002}]{tru02} Truss M.R., Wynn G.A., Murray J.R., King A.R., 2002, MNRAS,
337, 1329
\bibitem[\protect\citeauthoryear{Truss, Wynn \& Wheatley}{2004}]{tru04} Truss M.R., Wynn G.A., Wheatley P.J., 2004, 
MNRAS, 347, 569
\end{thebibliography}
\end{document}